\documentclass[twocolumn, secnumarabic, amssymb, nobibnotes, aps, prd, superscriptaddress, nofootinbib]{revtex4-2}

\usepackage[left=2cm, right=2cm, top=2.5cm, bottom=2.5cm]{geometry}

\newcommand{\tsb}{\textsubscript}

\usepackage{amsfonts}
\usepackage{amssymb}
\usepackage{amsmath}
\usepackage{dcolumn}
\usepackage{physics}
\usepackage{appendix}
\usepackage{theorem}

\usepackage{lipsum}
\usepackage{mathtools}

\usepackage{tabularx}
\usepackage{multirow}

\usepackage{hyperref}
\hypersetup{
    colorlinks=true,
    linkcolor=blue,     
    citecolor=blue,     
    urlcolor=blue,      
}

\usepackage[version=4]{mhchem}

\usepackage{pythonhighlight}

\usepackage{chemfig}
\usepackage{qcircuit}

\newcommand\arcbetweennodes[3]{%
\pgfmathanglebetweenpoints{\pgfpointanchor{#1}{center}}{\pgfpointanchor{#2}{center}}%
\let#3\pgfmathresult}
\newcommand\arclabel[6][black,-stealth,shorten <=1pt,shorten >=1pt]{%
\chemmove{%
\arcbetweennodes{#4}{#3}\anglestart \arcbetweennodes{#4}{#5}\angleend \draw[#1]([shift=(\anglestart:#2)]#4)arc(\anglestart:\angleend:#2); \pgfmathparse{(\anglestart+\angleend)/2}\let\anglestart\pgfmathresult \node[shift=(\anglestart:#2+1pt)#4,anchor=\anglestart+180,inner sep=0pt,
outer sep=0pt]at(#4){#6};}}

\let\oldAA\AA
\renewcommand{\AA}{\text{\normalfont\oldAA}}

\usepackage{graphicx} 
\usepackage{subcaption}
\usepackage{soul, xcolor}

\newcommand{\vtheta}{\boldsymbol{\theta}}

\newcommand{\avg}[1]{\langle #1 \rangle}

\usepackage{mdframed} 

\newmdenv[
  backgroundcolor=pink,
  linecolor=pink,
  skipabove=\topsep,
  skipbelow=\topsep,
  innertopmargin=0pt,
  innerbottommargin=0pt,
  leftmargin=0pt,
  rightmargin=0pt,
  innerleftmargin=0pt,
  innerrightmargin=0pt,
]{coloreddescription}

\usepackage[normalem]{ulem}
\usepackage{xcolor}

\begin{document}

\title{
    Improving VQE Parameter Quality on Noisy Quantum Processors with Cost-Effective Readout Error Mitigation
}

\author{Nacer Eddine Belaloui}
\email{nacer.belaloui@cqtech.org}
\affiliation{Constantine Quantum Technologies, \\Fr\`{e}res Mentouri University Constantine 1, Ain El Bey Road, Constantine, 25017, Algeria}
\affiliation{Laboratoire de Physique Math\'{e}matique et Subatomique, Fr\`{e}res Mentouri University Constantine 1, Ain El Bey Road, Constantine, 25017, Algeria}%
\author{Abdellah Tounsi}
\affiliation{Constantine Quantum Technologies, \\Fr\`{e}res Mentouri University Constantine 1, Ain El Bey Road, Constantine, 25017, Algeria}
\affiliation{Laboratoire de Physique Math\'{e}matique et Subatomique, Fr\`{e}res Mentouri University Constantine 1, Ain El Bey Road, Constantine, 25017, Algeria}%
\author{Abdelmouheymen Rabah Khamadja}
\affiliation{Constantine Quantum Technologies, \\Fr\`{e}res Mentouri University Constantine 1, Ain El Bey Road, Constantine, 25017, Algeria}
\affiliation{Laboratoire de Physique Math\'{e}matique et Subatomique, Fr\`{e}res Mentouri University Constantine 1, Ain El Bey Road, Constantine, 25017, Algeria}%
\author{Hamza Benkadour}
\affiliation{Laboratoire de Physique Math\'{e}matique et Subatomique, Fr\`{e}res Mentouri University Constantine 1, Ain El Bey Road, Constantine, 25017, Algeria}%
\author{Mohamed Messaoud Louamri}
\affiliation{Constantine Quantum Technologies, \\Fr\`{e}res Mentouri University Constantine 1, Ain El Bey Road, Constantine, 25017, Algeria}
\affiliation{Theoretical Physics Laboratory, University of
Science and Technology Houari Boumediene, BP 32 Bab Ezzouar,
Algiers, 16111, Algeria}
\author{Achour Benslama}
\affiliation{Constantine Quantum Technologies, \\Fr\`{e}res Mentouri University Constantine 1, Ain El Bey Road, Constantine, 25017, Algeria}
\affiliation{Laboratoire de Physique Math\'{e}matique et Subatomique, Fr\`{e}res Mentouri University Constantine 1, Ain El Bey Road, Constantine, 25017, Algeria}%
\author{Mohamed Taha Rouabah}
\email{m.taha.rouabah@umc.edu.dz}
\affiliation{Constantine Quantum Technologies, \\Fr\`{e}res Mentouri University Constantine 1, Ain El Bey Road, Constantine, 25017, Algeria}
\affiliation{Laboratoire de Physique Math\'{e}matique et Subatomique, Fr\`{e}res Mentouri University Constantine 1, Ain El Bey Road, Constantine, 25017, Algeria}%

\begin{abstract}
    The inherent noise in current Noisy Intermediate-Scale Quantum (NISQ) devices presents a major obstacle to the accurate implementation of quantum algorithms such as the Variational Quantum Eigensolver (VQE) for quantum chemistry applications. This study examines the impact of error mitigation strategies on VQE performance. We show that, for small molecular systems, an older-generation 5-qubit quantum processing unit (IBMQ Belem), when combined with optimized Twirled Readout Error Extinction (T-REx), achieves ground-state energy estimations an order of magnitude more accurate than those obtained from a more advanced 156-qubit device (IBM Fez) without error mitigation. Our findings demonstrate that T-REx, a computationally inexpensive error mitigation technique, substantially improves VQE accuracy not only in energy estimation, but more importantly in optimizing the variational parameters that characterize the molecular ground state. Consequently, state-vector simulated energies suggest that the accuracy of the optimized variational parameters provides a more reliable benchmark of VQE performance than quantum hardware energy estimates alone. Our results point to the critical role of error mitigation in extending the utility of noisy quantum hardware for molecular simulations.
\end{abstract}

\maketitle

\section{Introduction}

Quantum chemistry is a field that studies molecular systems and problems within the framework of quantum mechanics. Among these, the electronic structure problem is of central importance \cite{Szabo1996, Levine1999, helgaker2000}. In particular, determining the ground-state energy of a molecule is a fundamental task, as it relates to the most stable electronic configuration of the molecule and is essential to the prediction of molecular behavior for various chemistry applications such as in pharmaceutical research \cite{cao2018}, material science \cite{adjiman2021}, and biology \cite{fedorov2021}.

Classical methods that solve the ground-state energy problem vary in accuracy and computational cost. Full configuration interaction (FCI) methods yield exact solutions for a given level of theory but quickly become intractable even for relatively small systems \cite{lewars2010}. At the other extreme, the Hartree--Fock (HF) method \cite{Szabo1996} simplifies the problem by treating electron interactions in a mean-field approximation, reducing the electronic $n$-body problem to $n$ one-body problems at the cost of neglecting electron correlations. Ultimately, the most common methods attempt to balance computational accuracy and cost, but still face significant challenges for large or strongly correlated systems \cite{troyer2005, dral2020}.

Quantum methods, specifically quantum computing and quantum simulation, address the limitations of classical methods by natively leveraging quantum interactions \cite{benioff1980, feynman1982}. However, current quantum devices are still considered to be noisy intermediate-scale quantum (NISQ) era hardware \cite{preskill2018quantum}. This is when quantum noise and other coherent noise sources still impose practical limitations on the choice of quantum algorithms that one can implement.

In this context, variational quantum algorithms (VQAs) have emerged as promising tools for leveraging NISQ devices to obtain useful results \cite{Cerezo2021, Scriva2024}.
The variational quantum eigensolver (VQE) \cite{peruzzo2014variational}, an iterative, hybrid quantum-classical algorithm for approximating the minimum eigenvalue of a Hamiltonian, is one of the earliest VQAs to be introduced. The algorithm is suitable for the NISQ era due to its hybrid nature where the quantum computer's workload is kept low by delegating parts of the larger computation to a classical computer. Moreover, the noise resilience of the VQE can be improved through several techniques such as the inclusion of quantum error mitigation (QEM) \cite{Tilly2022} or the overparameterization of the ansatz \cite{Fontana2021}.
However, whether this noise resilience can be maintained at larger scales remains uncertain. Such experiments require more complex ans\"atze and introduce increased levels of noise.
Additional investigations on the impact of noise on the VQE's performance, particularly as quantum computers continue to evolve, are crucial to better understand this algorithm's relative noise resilience.\\

In this paper we show that a smaller and older-generation 5-qubit processor, when combined with error mitigation techniques, achieves similar to higher accuracies than a more advanced 156-qubit device for the same calculation without error mitigation \cite{Belaloui2025}. Specifically, we use the VQE to estimate the ground-state energy of \ce{BeH2} for a given molecular geometry. Despite \ce{BeH2} being a simple molecule, estimating its ground-state energy remains challenging for VQEs on early quantum devices with limited qubits and connectivity \cite{kandala2017, lolur2023, dobrautz2024}. We explore two types of ans\"atze: a hardware-efficient ansatz tailored for NISQ devices, and a physically informed ansatz designed to restrict the wave functions' search space to speed up the VQE. In this study, we employ the simultaneous perturbation stochastic approximation (SPSA) optimizer \cite{spall1998} for its relatively good ability to converge toward the ground state under noise within the VQE \cite{pellow2021}.
We analyze in this paper the VQE's performance through both ideal and noisy simulations, implement the algorithm on an IBM superconducting 5-qubit quantum device, and compare the results to those obtained on a more recent 156-qubit quantum device \cite{Belaloui2025}.\\

In the following, we begin by reviewing the general quantum chemistry problem at hand as well as the VQE in section \ref{sec:theory}. Then, in section \ref{sec:sim-qpu-results}, we present our simulations' fully detailed setup--both in the simulation and quantum hardware cases, the employed error mitigation strategy, and the simulation and quantum computer results. In section \ref{sec:discussion} we review and discuss the obtained results, and finally we conclude with section \ref{sec:conclusion}.

\section{Theory and Algorithm}
\label{sec:theory}

In this section, we briefly describe the problem's Hamiltonian and its mapping to a spin-operator form for use on a quantum computer. We also go over the principle behind the VQE and the two popular ans{\"a}tze that are used in this study.

\subsection{Electronic Hamiltonian}
\label{sec:quant_chem}

The electronic Hamiltonian describing the molecular electronic problem, when projected onto a discrete basis set, is given in its second quantized form in terms of fermionic annihilation and creation operators by
\begin{equation} \label{eq:2nd_quant}
	H_{el} =  \sum_{p,q} h_{pq} a^\dag_p a_q + \sum_{p,q,r,s} h_{pqrs} a^\dag_p a^\dag_q a_r a_s,
\end{equation}
where the first term represents the transition of single electrons between specific orbitals, while the second term corresponds to the mutual transition of electron pairs between specific orbitals. The coefficients $h_{pq}$ and $h_{pqrs}$ are the one- and two-electron integrals defined as \cite{helgaker2008quantitative, lewars2010}
\begin{align}
    \label{eq:h_pq}
    h_{pq} 
    &= \int \psi^*_p(\mathbf{x})\left(\frac{-\nabla^2}{2} - \sum_{i} \frac{Z_i}{|\mathbf{R}_i - \mathbf{r}|}\right)\psi_q(\mathbf{x}) d\mathbf{x},\\
    \label{eq:h_pqrs}
    h_{pqrs} &= \int \frac{\psi^*_p(\mathbf{x}_1)  \psi^*_q(\mathbf{x}_2)  \psi_r(\mathbf{x}_1)  \psi_s(\mathbf{x}_2)}{|\mathbf{r}_1 - \mathbf{r}_2|} d\mathbf{x}_1 d\mathbf{x}_2,
\end{align}
where $\mathbf{R}_i$ and $\mathbf{r}$ are the nucleus and electron positions, and $\psi(\mathbf{x})$ is the molecular spin orbital's wave function with $\mathbf{x}$ describing both the electron's position and spin.
The Hamiltonian in this form requires a finite number of qubits and is easily mapped to quantum gates.
A detailed derivation of equations \eqref{eq:2nd_quant}, \eqref{eq:h_pq}, and \eqref{eq:h_pqrs}, as well as further quantum chemistry details for the electronic problem are provided in \citet{Belaloui2025}.

\subsection{Fermionic to Spin Operators Mapping}
\label{sec:mapping}
On quantum computers, we mainly operate on qubits with Pauli operators. However, a transition from eq. \eqref{eq:2nd_quant} to its equivalent in Pauli operators must maintain the fermionic anti-commutation relations that describe the creation and annihilation operators. Several methods have been developed to take this requirement into account. Among the most popular we mention the Jordan-Wigner \cite{Jordan1928}, Bravyi-Kitaev \cite{bravyi2002}, and parity \cite{Seeley2012} transformations. 
In this work, we use the parity mapping with qubit tapering since it allows for quantum computational resource reduction and provides, in this case, lighter Pauli terms that require fewer local measurements.\\

In our implementation for the \ce{BeH2} case, the number of required qubits and Pauli terms is dictated by the amount of approximation introduced in limiting the number of active orbitals and electrons to reduce computational costs. A heavy approximation is not generally recommended but can be necessary for small quantum devices that do not possess the required number of qubits for more accurate and larger Hamiltonians. Qubit tapering through parity mapping, however, reduces the number of required qubits without affecting the Hamiltonian's accuracy. This tapering is a result of fixing the number of electrons in each of the up and down spin sectors \cite{Belaloui2025}.
In this study, we consider 3 active orbitals, which translates into a requirement of 4 qubit instead of 6 thanks to the qubit tapering that is enabled by the parity mapping. This enabled us to simulate 3 active molecular orbitals on the 5-qubit IBMQ Belem quantum computer \cite{IBMQ2024}.

\subsection{The Variational Quantum Eigensolver}
\label{sec:vqe}

\begin{figure}[t]
    \centering
    \includegraphics[width=0.95\linewidth]{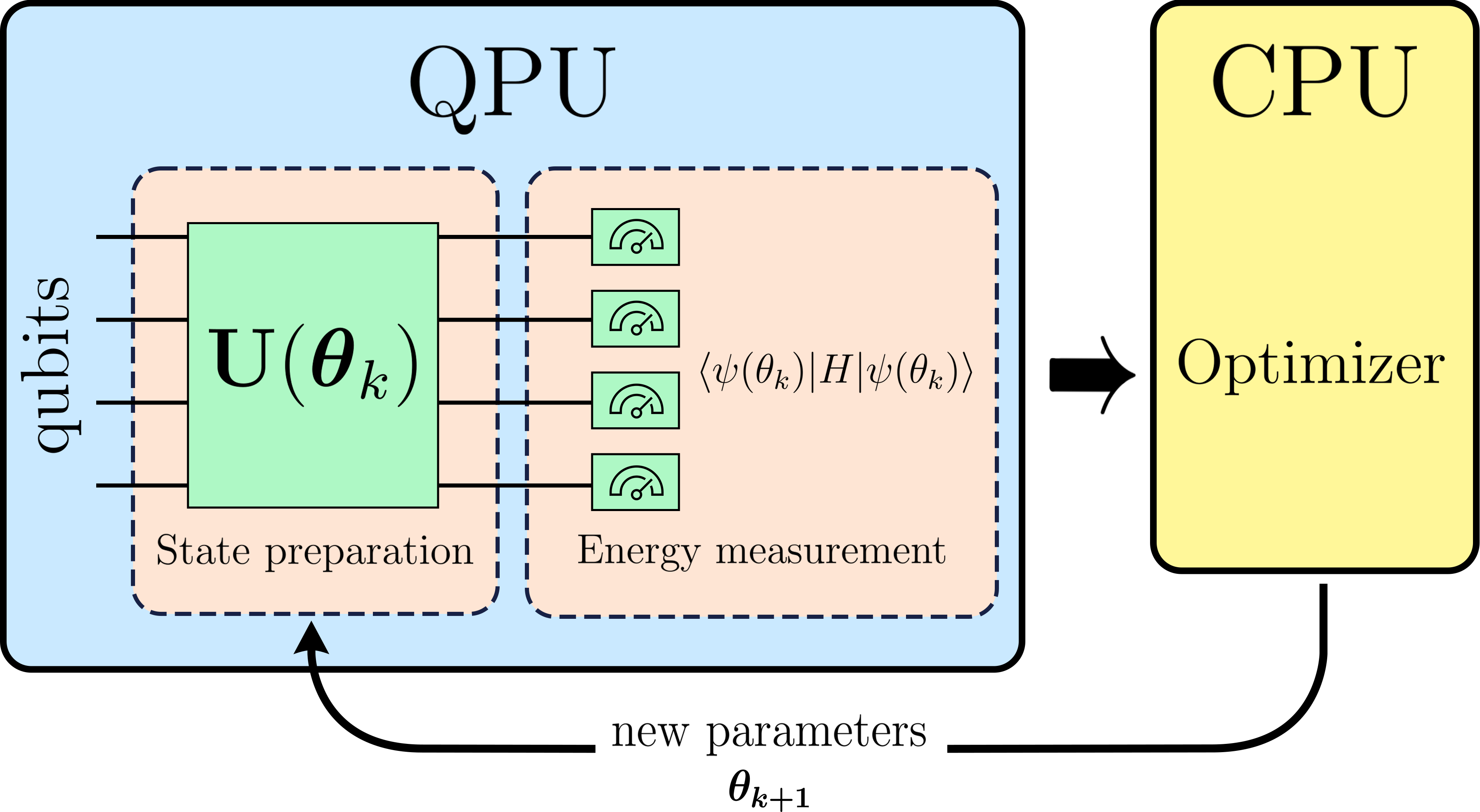}
    \caption{The variational quantum eigensolver's optimization loop. At each iteration $k$, a quantum state is first prepared using a set of parameters $\vtheta_k$, then its energy is measured on the QPU. The CPU takes in the QPU measurements and produces the next set of parameters, $\vtheta_{k+1}$, following a classical optimization algorithm.}
    \label{fig:vqe-diagram}
\end{figure}

The Variational Quantum Eigensolver (VQE) is a hybrid quantum-classical algorithm initially designed to approximate the ground state of a quantum mechanical system \cite{peruzzo2014variational}; the state with the minimum energy. In quantum chemistry, the VQE is usually used to solve the electronic structure problem of a molecular system that can be described by the Hamiltonian in eq. \eqref{eq:2nd_quant}. This is achieved by minimizing a quantity, usually the energy, associated with a given trial state $\psi$ that is parameterized with a set of free parameters $\vtheta$. Explicitly, the VQE seeks to minimize
\begin{equation}
    E(\vtheta) = \bra{\psi(\vtheta)}H \ket{\psi(\vtheta)},
    \label{eq:energy_theta}
\end{equation}
for a normalized state $\ket{\psi(\vtheta)}$, with the Rayleigh-Ritz theorem \cite{zettili2009, Tannoudji2017-tome3} ensuring that for a given Hamiltonian $H$, any $\ket{\psi(\vtheta)}$ yields
\begin{equation}
    E(\vtheta) \ge E_0.
    \label{eq:ritz}
\end{equation}
That is, $E_0$ is a lower bound on the VQE result. Fig. \ref{fig:vqe-diagram} shows a simplified diagram that describes the VQE's optimization loop.\\

Choosing an appropriate ansatz is a key step in the implementation of VQE, as it will eventually dictate the accuracy of the computations. There exist multiple families of ans\"atze, though, two types are most common in quantum chemistry applications: the \textit{hardware-efficient ans\"atze} (HEAs) \cite{kandala2017, Sim2019, Tilly2022} that are designed around the quantum hardware limitations, and the chemically-inspired ans\"atze, notably the \textit{Unitary Coupled-Cluster} (UCC) ansatz class \cite{watts1989, lee2018generalized}, which are built following chemical and physical motivations. HEAs usually consist of multiple alternating layers of single qubit rotations followed by entangling layers that tend to largely respect the hardware's native connectivity. This class of ans\"atze is aimed to keep hardware-induced errors to a minimum by reducing the depth of the transpiled circuit, at the cost of being less trainable and less accurate in approximating the ground state. By contrast, chemically-inspired ans\"atze are more expressive since they model the problem's dynamics. The UCC ansatz class models the excitations of electrons from occupied orbitals to virtual unoccupied ones. Naturally, these ans\"atze are more accurate in ideal conditions but are heavily affected by the high noise present in current quantum computers due to the increased depth of their transpiled quantum circuits.\\

Moreover, efforts in exploring the intersection of these two classes resulted in several interesting ans\"atze that aim to be both hardware efficient while retaining a chemical motivation. These include the \emph{$k$-Unitary pair Coupled Cluster Generalized Singles and Doubles} ansatz \cite{lee2018generalized} that has a significant reduction in depth over the conventional UCCSD, and the \emph{Local Unitary Cluster Jastrow} ansatz \cite{motta2023bridging} that produces a circuit without SWAP gates, which significantly reduces its depth and susceptibility to noise. These two mentioned ans\"atze are aimed at large-scale applications and are not intended for smaller systems such as the one treated in this paper \cite{lee2018generalized, motta2023bridging}.\\

In this work we use the \textit{Efficient SU2} ansatz as our HEA of choice, and the \emph{Unitary Coupled Cluster Singles and Doubles} (UCCSD) ansatz as the chemically-inspired ansatz, for comparison purposes, where we only consider single- and double-electron excitations \cite{grimsley2022, bakoutsos2018, Belaloui2025}.\\

In addition to a good ansatz, an appropriate optimizer must be chosen. This choice will significantly impact the VQE's overall cost and performance \cite{pellow2021}. We chose for this study the \textit{Simultaneous Perturbation Stochastic Approximation (SPSA)} algorithm as our optimizer.
Originally developed to optimize fluctuating, non-deterministic cost functions in classical applications \cite{spall1998}, SPSA is now widely used in quantum computing, notably within variational quantum algorithms \cite{pellow2021}.

\section{Experiments and Results}
\label{sec:sim-qpu-results}
In this section, we apply the VQE algorithm for the BeH\tsb{2} molecule's ground-state energy problem. Specifically, we simulate the VQE in the ideal and noisy cases.
In both cases, we use both the chemically-inspired \textit{UCCSD} and the hardware-efficient \textit{Efficient SU2} ans\"atze. We also implement the VQE on the now retired IBMQ Belem quantum processing unit (QPU), but only using the \textit{Efficient SU2} ansatz. As IBMQ Belem was a 5-qubit QPU, we restricted ourselves to an active space approximation considering $2$ electrons and $3$ active molecular orbitals. Such a system has an exact ground-state energy
\begin{equation}
    E_0^{\text{target}} = -15.56089 \text{ Ha}. 
    \label{eq:ref_energy}
\end{equation}
We keep the same experimental setup for both the simulation and QPU experiments, with the addition of an error-mitigation setup during the latter. The following subsections' results are summarized in Table \ref{table:results} below.

\subsection{Simulations}
\label{subsec:sim}

\begin{figure*}[ht!]
    \centering
    \begin{subfigure}[b]{0.49\linewidth}
        \centering
        \includegraphics[width=\linewidth]{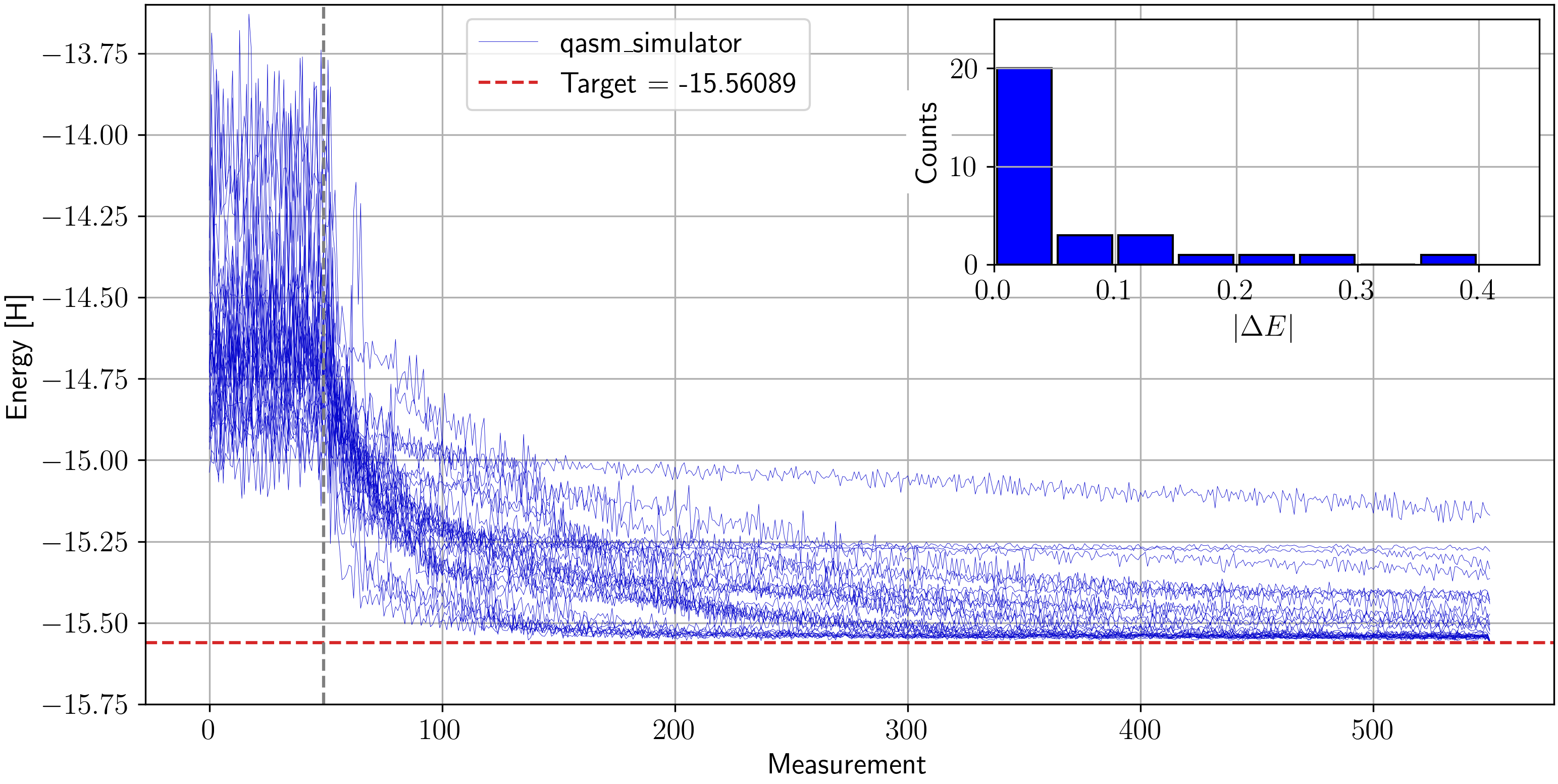}
        \caption{30 ideal VQEs with the \textit{Efficient SU2} ansatz. $\langle E \rangle_\text{best 5} = -15.56043 \text{ Ha}$, $\sigma=0.00008$.}
        \label{fig:effsu2_svs}
    \end{subfigure}
    \hfill
    \begin{subfigure}[b]{0.49\linewidth}
        \centering
        \includegraphics[width=\linewidth]{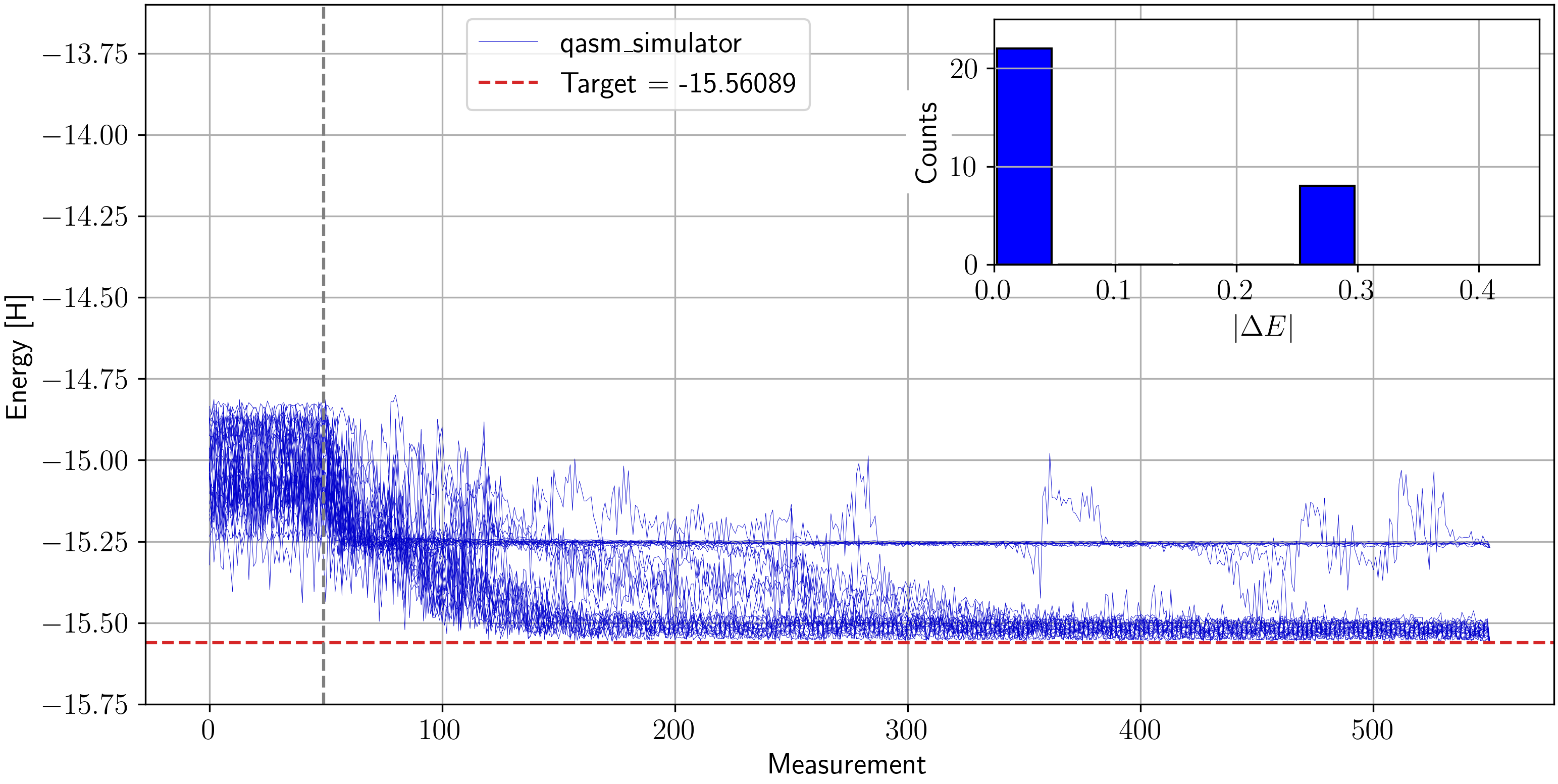}
        \caption{30 ideal VQEs with the \textit{UCCSD} ansatz. $\langle E \rangle_\text{best 5} = -15.56077 \text{ Ha}$, $\sigma=0.00003$.}
        \label{fig:uccsd_svs}
    \end{subfigure}
    
    \vspace{0.5cm} 
    
    \begin{subfigure}[b]{0.49\linewidth}
        \centering
        \includegraphics[width=\linewidth]{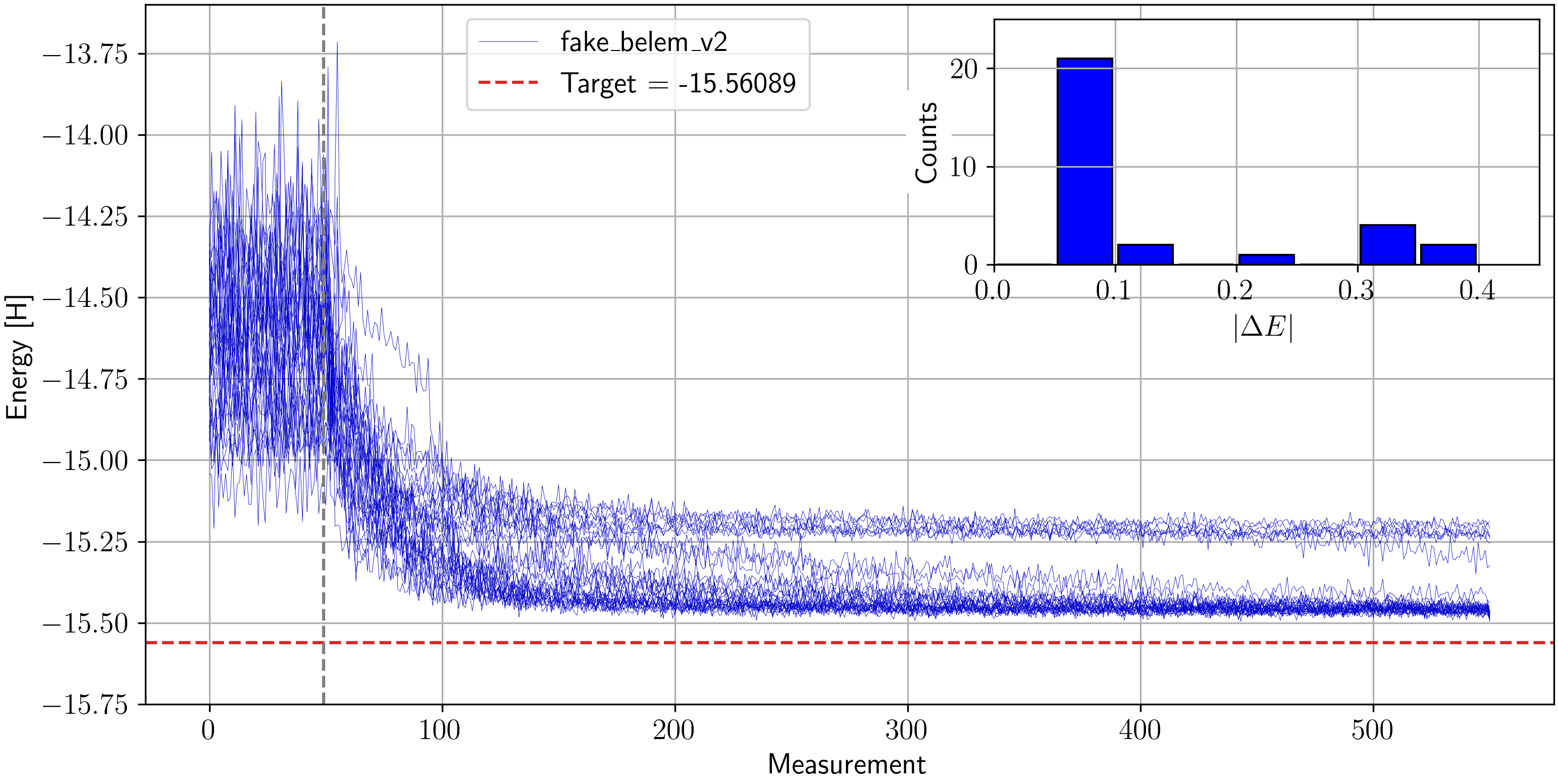}
        \caption{30 noisy VQEs with the \textit{Efficient SU2} ansatz. $\langle E \rangle_\text{best 5} = -15.49035 \text{ Ha}$, $\sigma=0.00404$.}
        \label{fig:effsu2_noisy}
    \end{subfigure}
    \hfill
    \begin{subfigure}[b]{0.49\linewidth}
        \centering
        \includegraphics[width=\linewidth]{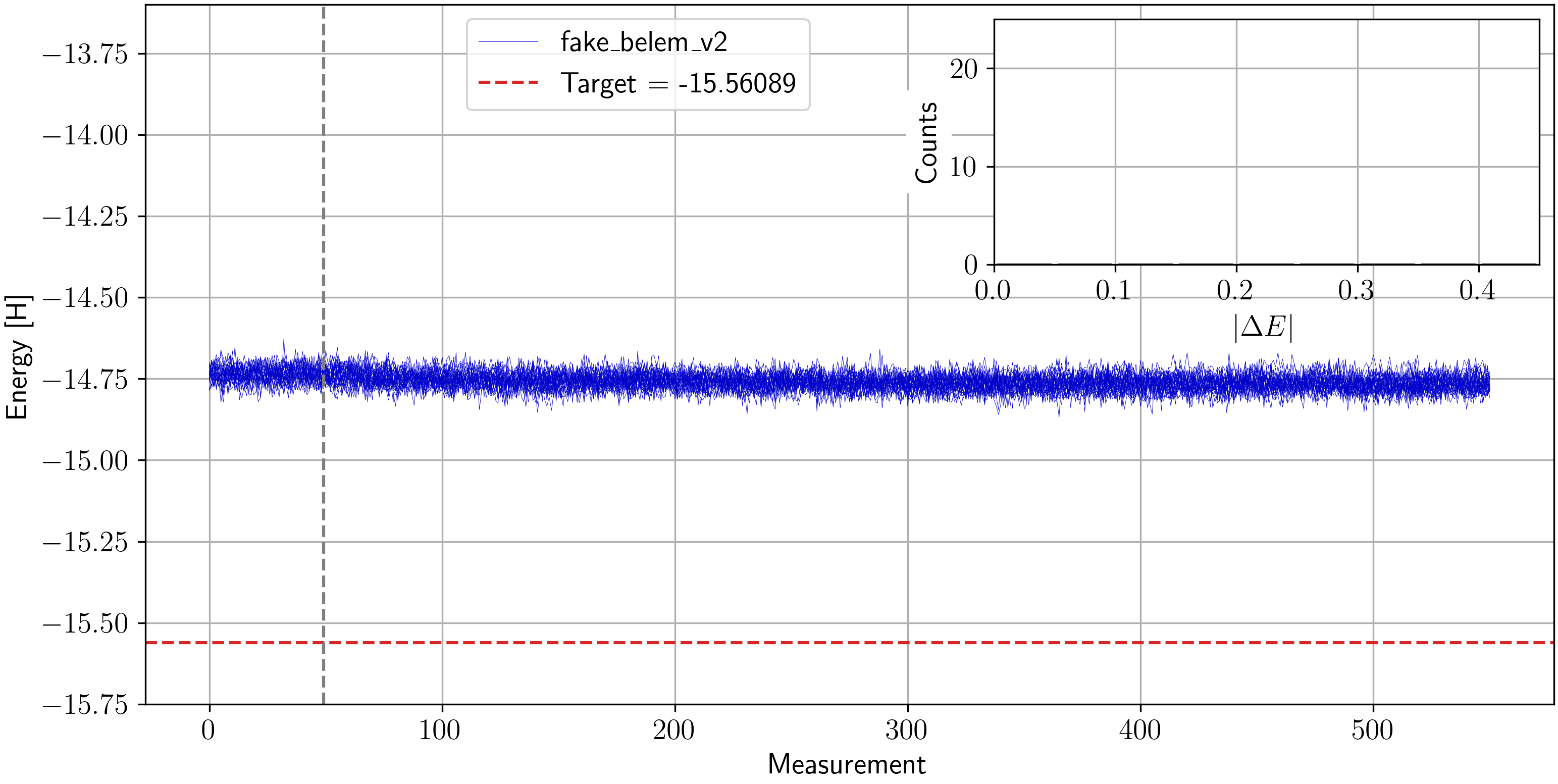}
        \caption{30 noisy VQEs with the \textit{UCCSD} ansatz. $\langle E \rangle_\text{best 5} = -14.7918 \text{ Ha}$, $\sigma=0.0086$.}
        \label{fig:uccsd_noisy}
    \end{subfigure}

    \caption{
    Comparison of the \textit{Efficient SU2} and \textit{UCCSD} ans\"atze under ideal and noisy conditions (using the IBMQ Belem noise model). In each figure are 30 VQE convergence graphs, each containing both the SPSA calibration measurements and the VQE measurements. The histograms show how many VQEs converged to under a given $|\Delta E|$ value, with increasing steps of $0.05 \text{ Ha}$. The average of the best 5 values, $\langle E \rangle_\text{best 5}$, and the standard deviation, $\sigma$, are given in each case.
    For reference, the Hartree--Fock energy is $E_\textit{\rm HF} = -15.56033 \text{ Ha}$.
    }
    \label{fig:all_simulations}
\end{figure*}

In the following ideal and noisy simulations, we will be using the SPSA optimizer with the following set of hyperparameters $\alpha = 0.602$, $\gamma = 0.101$, $A = 0$, and $c=0.2$ \cite{spall2003, kandala2017}. The perturbation series $c_k$ for the SPSA optimizer is determined by the chosen hyperparameters, and the learning rate series $a_k$ is calibrated with $50$ calls to the cost function before the optimization starts. We opt to stop the optimization after $250$ iterations, thus, performing a total of $551$ measurements: $50$ for calibration, $2 \times 250$ for the gradients estimation at each iteration to optimize the parameters, and $1$ measurement of the final energy which the algorithm converges to. Each of the performed measurements is taking $4000$ shots, meaning it is measuring each quantum circuit $4000$ times and computing the energy from the resulting distribution of results obtained from the measurements.

\subsubsection{Ideal case}

The first set of simulations are performed under ideal noise conditions, that is, no quantum noise is simulated in quantum circuits. Only fluctuations from quantum measurements, commonly known as shot noise, are present. These noise-free simulations constitute an upper bound for how the VQE would ideally perform.

The subfigures in Fig. \ref{fig:all_simulations} show the progression of the simulated VQEs using both the \textit{Efficient SU2} and \textit{UCCSD} ans\"atze, starting with randomized parameters, and using the SPSA algorithm for parameter optimization. In subfigures \ref{fig:effsu2_svs} and \ref{fig:uccsd_svs}, we observe that both ans\"atze ended up converging towards the target $E_0^{\text{target}}$ within less than $1$ millihartree when taking the average over the best $5$ results, with the \textit{UCCSD} ansatz performing better. Moreover, we notice a faster convergence for the \textit{UCCSD} compared to \textit{Efficient SU2}, but with a number of VQEs ending up in a local minimum situated at around $-15.25 \text{ Ha}$.

\subsubsection{Noisy simulations}

\begin{figure*}[ht!]
    \centering
    \includegraphics[width=0.95\linewidth]{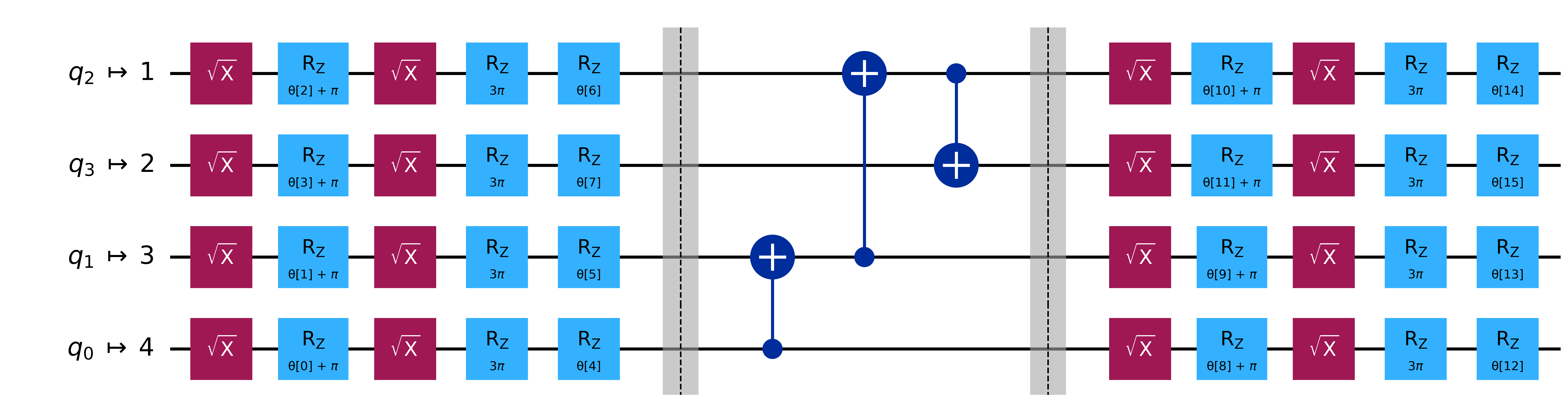}
    \caption{
        The transpiled \textit{Efficient SU2} circuit that runs on IBMQ Belem. This circuit only uses gates that are native to the QPU, in this case the $\operatorname{\sqrt{X}}$, $\operatorname{R_Z}$, and $\operatorname{CNOT}$ gates. Therefore, the $\operatorname{R_Y}$ gates have been decomposed, while the $\operatorname{CNOT}$ gates have been kept but rearranged to follow the QPU's connectivity. On the left of the circuit, the mapping between the logical and physical qubits is shown.
    }
    \label{fig:transpiled_effsu2_circuit}
\end{figure*}

The second set of simulations includes the simulation of several of the IBMQ Belem device characteristics, including its levels of quantum noise. This noisy simulation also changes the actual quantum circuit that we run, as additional qubit connectivity and basis quantum gate constraints are imposed on the final transpiled circuit to match the real QPU's design. We show the final transpiled circuit in Fig. \ref{fig:transpiled_effsu2_circuit}.

Similarly to the ideal simulations, for each of the \textit{Efficient SU2} and \textit{UCCSD} ans\"atze, we carried out a set of 30 VQE runs. The results for both sets are respectively aggregated in figures \ref{fig:effsu2_noisy} and \ref{fig:uccsd_noisy}. The two sets of simulations confirmed that the hardware-efficient ansatz is far better suited for the forthcoming real QPU experiment. Indeed, the average over the 5 best energy results from the \textit{Efficient SU2} set gives
$$\langle E_\text{EffSU2}^\text{noisy} \rangle_\text{best 5} = -15.49035 \text{ Ha,}$$
while the 5 best \textit{UCCSD} energies give
$$\langle E_\text{UCCSD}^\text{noisy} \rangle_\text{best 5} = -14.79180 \text{ Ha.}$$
We have also classically evaluated, through a noiseless state-vector simulator (SVS), the energies corresponding to the $\vtheta^\text{noisy}$ parameter vectors resulting from the 5 best-performing noisy VQEs.
For the \textit{Efficient SU2} ansatz, this average is
$$ \langle E_{\rm EffSU2}^\text{SVS}(\vtheta^\text{noisy}) \rangle_\text{best 5} = -15.55917 \text{ Ha,}$$
whereas for the \textit{UCCSD} case we obtain
$$ \langle E_{\rm UCCSD}^\text{SVS}(\vtheta^\text{noisy}) \rangle_\text{best 5} = -15.24008 \text{ Ha.}$$
This confirms that the \textit{UCCSD} VQEs did not converge at all, contrary to the \textit{Efficient SU2}.

\subsection{Real hardware}

\begin{figure}[t!]
    \includegraphics[width=0.95\linewidth]{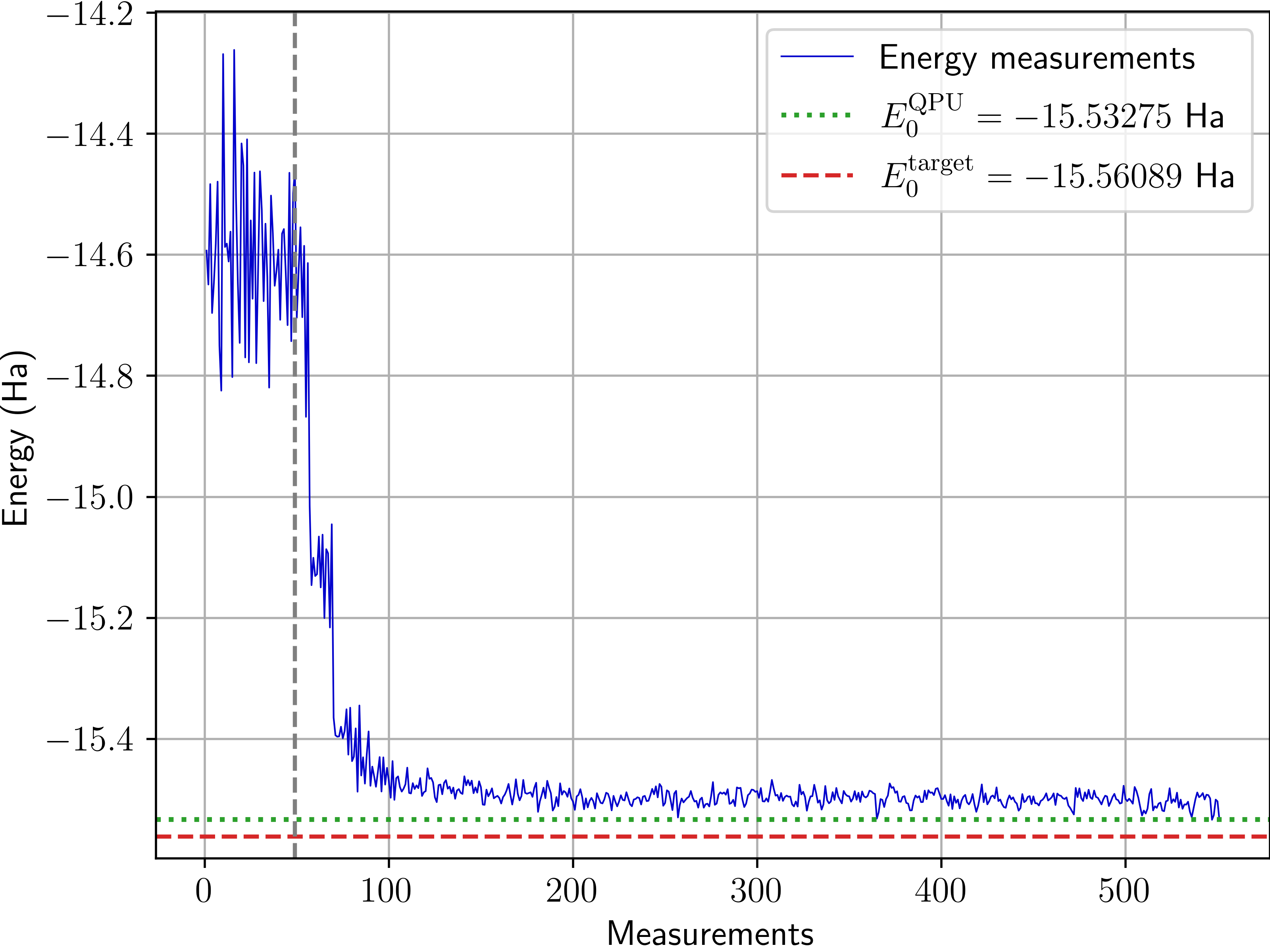}
    \centering
    \caption{
        All the VQE measurements performed on IBMQ Belem. This includes SPSA's $50$ calibration measurements (left of the gray cutoff) and $2$ measurements per iteration over a total of $250$ iterations. One final energy measurement is performed at the end (green dotted line).
    }
	\label{fig:belem-graph}
\end{figure}

\begin{figure}[t!]
	\includegraphics[width=0.95\linewidth]{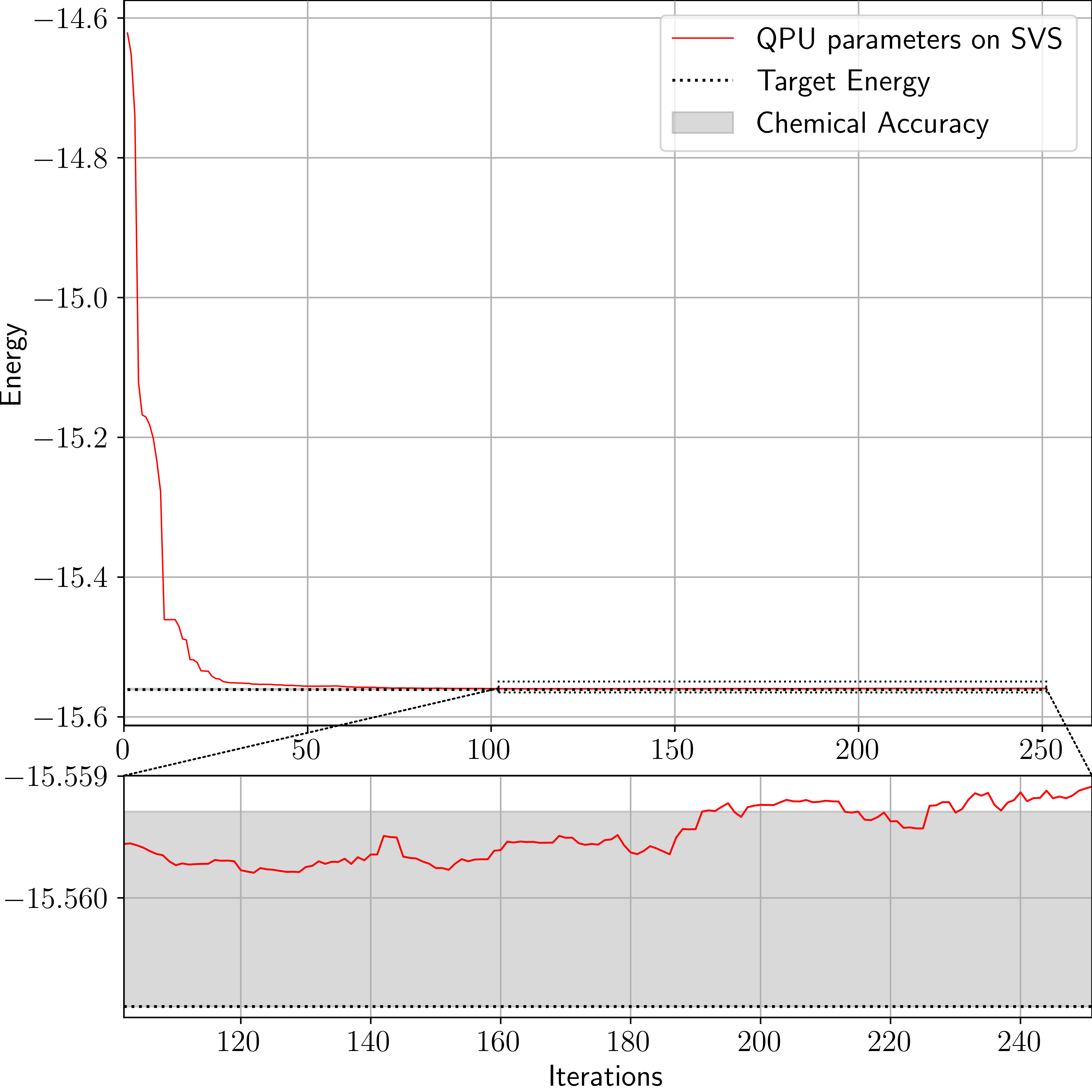}
	\centering
	\caption{
            SVS-evaluated energies at each iteration of the VQE. The average energy over the last 10\% of iterations $\langle E^\text{SVS}(\vtheta^\text{QPU}) \rangle$ is $-15.55918(5)$ Ha, with a difference $|\Delta E| = 0.00171$ Ha from the target energy. The final measurement is reported separately.
        }
	\label{fig:belem-qpu-on-svs}
\end{figure}

We performed a VQE experiment for the \ce{BeH2} molecule on the (now retired) 5-qubit IBMQ Belem QPU, using the SPSA algorithm as the classical optimizer, as well as the twirled readout error extinction (T-REx) \cite{Ewout2022} technique---which we will describe in full below---for error mitigation. As with the simulated experiments, the optimization was carried out over $250$ iterations, with each iteration requiring $2$ energy measurements to estimate the gradient.

In terms of quantum computational requirements, each energy measurement involved evaluating $7$ qubit-wise commuting Pauli groups, that is, $7$ circuits are executed for each energy measurement. As part of the utilized error mitigation scheme, these groups were themselves measured using $16$ twirled circuits each, with $250$ shots per circuit, resulting in $4000$ shots per group and $2 \times 7 = 14$ measurements per iteration. Additionally, T-REx required an additional overhead of $8192$ shots. Each iteration thus required the execution of $240$ circuits totaling $64192$ shots.

Following optimization, a final energy estimation was performed using the optimized ansatz parameters. In total, $551$ QPU measurements were executed, including those from SPSA's calibration phase. The complete timeline of QPU measurements is shown in Fig.\ref{fig:belem-graph}. The total QPU execution time was $7667$ seconds (i.e., $127$ minutes and $47$ seconds). In Table \ref{tab:vqe-config}, we present the detailed settings used at each single VQE iteration on the QPU. Additionally, using the final VQE parameters, we perform 3 zero-noise extrapolation measurements on IBMQ Belem, as will be further explained below.\\

\begin{table*}[t]
    \centering
    \begin{tabular}{p{0.30\textwidth} p{0.45\textwidth}}
        \hline
        \textbf{Setting} & \textbf{Value} \\[0.5ex]
        \hline
        \textbf{Measurements per iteration}  & $2$ (gradient only) \\
        \textbf{Shots per single measurement}       & $4000$ shots \\
        \textbf{Measurement circuits}           & $7$ circuits (Qubit-wise commuting Pauli groups) \\
        \textbf{Twirled circuits}            & $16$ twirled circuits, $250$ shots per twirled circuit \\
        \textbf{T-REx calibration circuits}    & $16$ circuits (4 qubits $\rightarrow 16$ states)\\
        \textbf{Total calibration shots}       & $8192$ shots \\
        \textbf{Total shots per iteration}   & $2 \times 7 \times 4000 + 8192 = 64192$ shots \\
        \textbf{Total circuits without calibration} & $2 \times 7 \times 16 = 224$ circuits \\
        \textbf{Total circuits}              & $240$ \\
        \hline
    \end{tabular}
    
    \caption{The detailed per-iteration measurement scheme used in the IBMQ Belem VQE implementation. This includes both the energy estimation and T-REx mitigation required resources.}
    \label{tab:vqe-config}
\end{table*}

The converged ground-state energy as measured at the end of the QPU-implemented VQE experiment is
\begin{equation}
    E_0^\text{QPU} = -15.53275 \text{ Ha}.
\end{equation}
This value having been obtained over $4000$ shots, it is also associated with a measurement standard deviation of
\begin{equation}
    \sigma^\text{QPU} = 0.46579 \text{ Ha}.
\end{equation}
We also report the final optimized parameters $\vtheta^\text{QPU}$ to be
\begin{equation}
    \boldsymbol{\theta}^\text{QPU} = 
    \left(
    \begin{array}{lcccc}
              &\text{Ry} & \text{Rz} & \text{Ry} & \text{Rz} \\
    \text{Q}_1 & +3.255  & -2.208 & -0.033 & +0.702 \\
    \text{Q}_2 & -0.056  & +1.688 & +0.058 & -0.273 \\
    \text{Q}_3 & +3.222  & -0.416 & -0.038 & -0.137 \\
    \text{Q}_4 & -2.214  & +0.113 & +2.204 & +1.747
    \end{array}
    \right),
\end{equation}
where we have explicitly shown each parameter with its associated quantum gate and qubit.

Energy estimation using the final QPU-optimized parameter vector $\boldsymbol{\theta}^\text{QPU}$ on a perfect SVS yields
\begin{equation}
    E_0^\text{SVS}(\boldsymbol{\theta}^\text{QPU}) = - 15.55539 \text{ Ha,}
\end{equation}
$22.64$ mHa lower than the QPU-estimated $E_0^\text{QPU}$. This shows that despite the relatively high levels of noise of the IBMQ Belem device, and the imperfect raw energy values obtained from the QPU measurements, the VQE was still able to optimize the ansatz parameters to some extent. Furthermore, when re-evaluating all the produced parameters, $\vtheta^\text{QPU}$, we find that the best set of parameters was produced at iteration $121$, resulting in

\begin{equation}
    E^{\rm SVS}(\vtheta_{\rm 121}^{\rm QPU}) = -15.55979\text{ Ha.}
\end{equation}

\subsection{Error mitigation}
Despite the VQE's ability to converge in general towards approximate ground state energies with imperfect quantum devices, the presence of noise and gate errors can still significantly affect the accuracy of the results. Error mitigation \cite{Tilly2022, endo2021} plays a crucial role in the successful implementation of the VQE, particularly as quantum hardware is still in the NISQ era.  Error mitigation techniques help in reducing the impact of these errors without the need for full quantum error correction, which is yet to be demonstrated for useful computations \cite{Quantinuum2024, Google2025}. By incorporating these strategies, we aim to achieve more reliable and accurate energy estimates in the NISQ era, making the VQE a more robust tool for quantum chemistry applications, especially as we scale to more complex molecular systems.\\

\subsubsection*{Readout Error Mitigation}
In our implementation of the VQE on IBMQ Belem, we have employed a readout error mitigation (REM) technique during the algorithms iterative process, that is, only aiming at reducing readout errors from the quantum computer. The most straightforward way to mitigate this kind of errors is a simple brute-force approach to REM, where we assume that there exists a classical noise map, defined by a matrix $A$, which encodes the probabilities that a measurement output bitstring $x$ is measured as another bitstring $y$ \cite{Bravyi2021}. A simple one-qubit example is given as
$$
\begin{array}{c@{\quad}c}
    \begin{array}{c}
        \\
        A =
        \\ 
    \end{array} &
    \begin{array}{c@{\quad}c}
        & 
        \begin{array}{cccc}
            0 & & & 1
        \end{array} \\
        \begin{array}{c}
            0 \\
            1
        \end{array} &
        \left(
        \begin{array}{cc}
            1 - \epsilon_0 & \epsilon_1 \\
            \epsilon_0 & 1 - \epsilon_1
        \end{array}
        \right),
    \end{array}
\end{array}
$$
where $1-\epsilon_0$ and $1-\epsilon_1$ are the classical probabilities of measuring the correct bits $0$ and $1$, respectively, and $\epsilon_0$ and $\epsilon_1$ define readout error rates. One might then simply apply the inverse matrix $A^{-1}$ to the vector of frequencies of the measured output states for a given quantum circuit, or in the previous example, the states $\ket{0}$ and $\ket{1}$. This, of course, although effective at small numbers of qubits, does not scale efficiently, as the matrix $A$ grows exponentially with the number of qubits.\\

\subsubsection*{Twirled Readout Error extinction}
The twirled readout error extinction (T-REx) method, as the name implies, is also a readout error mitigation method, but which improves upon the previously described brute-force method \cite{Bravyi2021, Ewout2022}. T-REx uses Pauli twirling to transform the original noise map, $A$, into a twirled noise map, $A^*$, that is diagonal in the $Z$-basis, such that after twirling, its effect is measured only as a multiplicative constant $\lambda$ acting on the measured expectation value. The T-REx process thus primarily aims at measuring the $\lambda$ constant. The global scheme for applying the T-REx workflow on a given quantum circuit $C$, as described in \citet{Ewout2022}, is presented here in its main steps:
\begin{enumerate}
    \item Calibration: estimate the multiplicative factor $\lambda$ by running the identity circuit with random Pauli twirls ($X$ gates).
    \item Measurement: run the quantum circuit $C$, appended with random Pauli twirls, to obtain an initial noisy expectation value.
    \item Divide out $\lambda$ from the noisy expectation value obtained in step 2, and produce a T-REx-mitigated expectation value.
\end{enumerate}

The aforementioned Pauli twirling is an error mitigation technique in itself \cite{Knill2008, Li2017}, where random Pauli gates are inserted before and after noisy operations in a quantum circuit, producing a set of \textit{``twirled circuits"} whose results are then averaged over to obtain a mitigated result. However, in the context of T-REx, random Pauli bitflips are only applied before the measurement. That is, we do not use Pauli twirling as a means to mitigate errors in itself.

Although T-REx appears as a computationally cheap error mitigation technique, one must take into account that it has the drawback of requiring a calibration measurement overhead, which can however be kept reasonably small in the case of weak noise \cite{Ewout2022}. On the other hand, it has the benefit of greatly improving upon the brute-force REM method, as the noise map $A$ (and consequently $A^*$) is not directly computed or explicitly dealt with.

\subsubsection*{Zero-Noise Extrapolation}
Zero-noise extrapolation (ZNE) \cite{temme2017} is a powerful error mitigation technique used in quantum computing to enhance the accuracy of computations on noisy quantum devices. The core idea behind ZNE is to systematically increase the noise in a controlled manner during the quantum computation, then use the results from these noisier computations to extrapolate back to what the result would have been in the absence of noise. This method is well suited for current quantum devices due to the simplicity of its implementation. By effectively reducing the noise effects on measured expectation values, ZNE can significantly improve the reliability of quantum algorithms such as the VQE, bringing us closer to achieving accurate quantum simulations and computations despite the limitations of existing quantum hardware. However, in the case of the VQE, it may be best used to enhance the accuracy of the final result rather than during the optimization to keep the algorithms total runtime low.

\begin{figure*}[t!]
    \centering
    \begin{subfigure}[b]{0.49\linewidth}
        \centering
        \includegraphics[width=\linewidth]{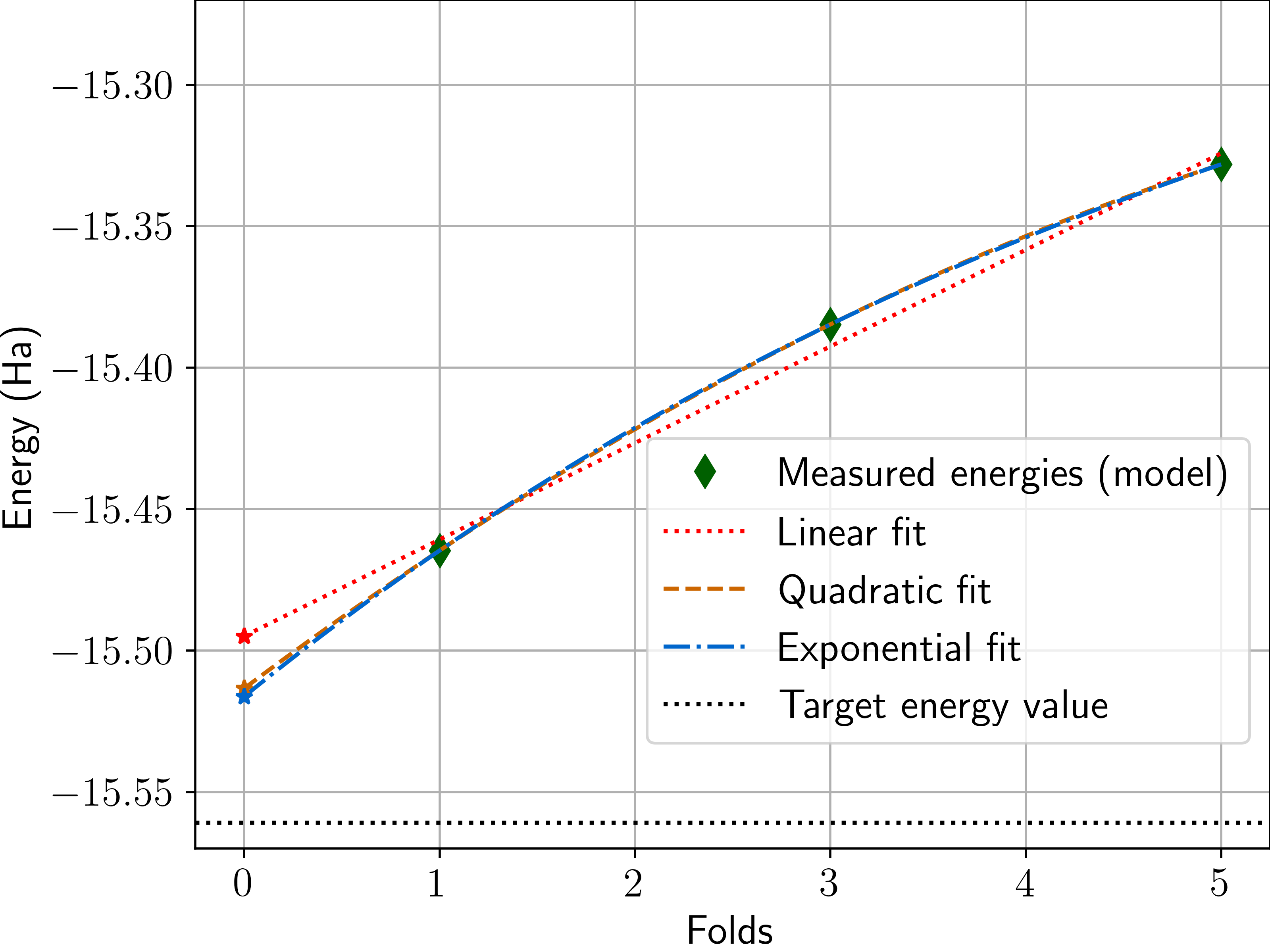}
        \caption{
            ZNE on IBMQ Belem's noise model. The mitigated values obtained with the linear, quadratic, and exponential fittings were $-15.49489$, $-15.51337$, and $-15.51624$ Ha, respectively.
        }
        \label{fig:zne_effsu2_model}
    \end{subfigure}
    \hfill
    \begin{subfigure}[b]{0.49\linewidth}
        \centering
        \includegraphics[width=\linewidth]{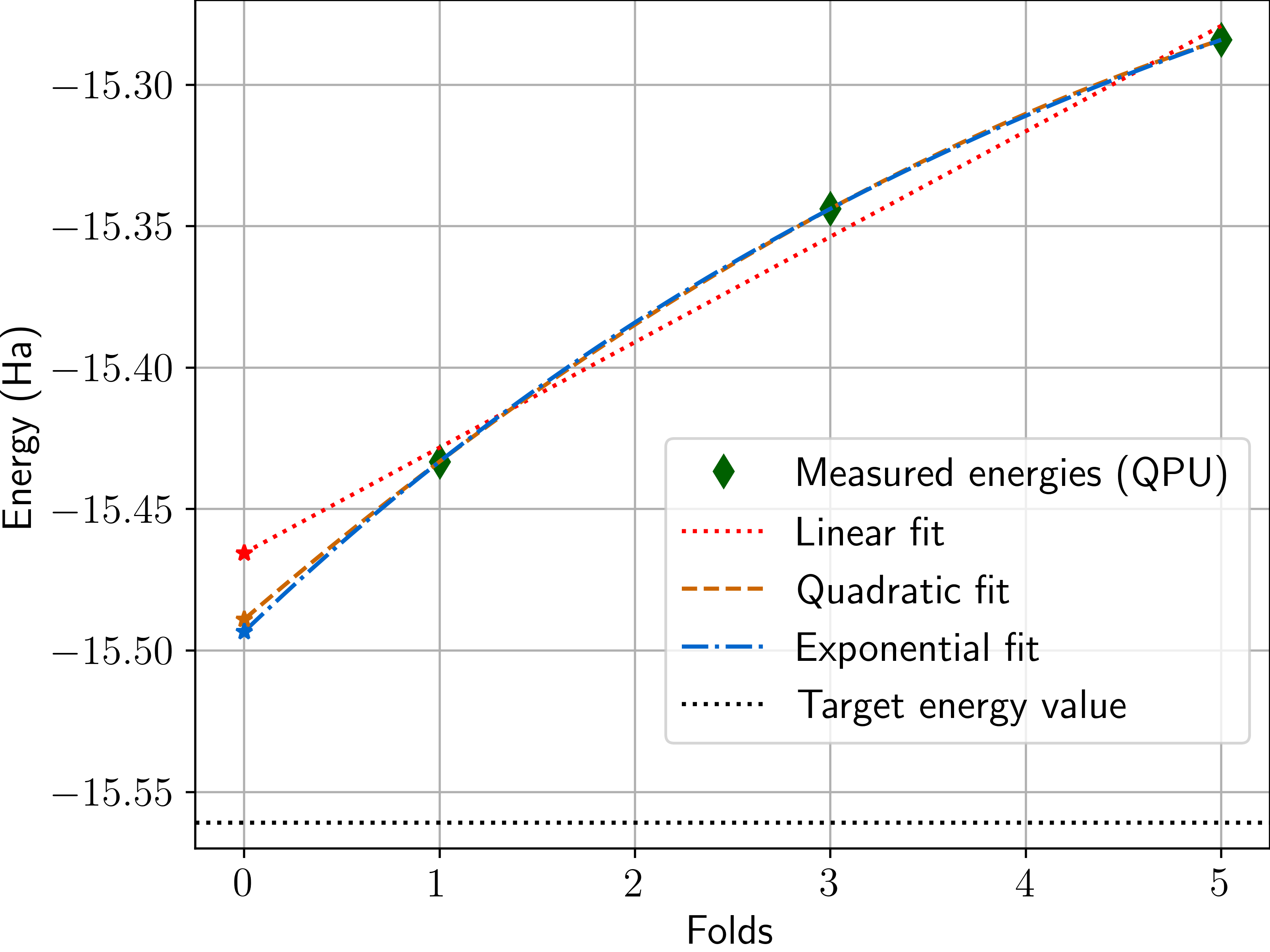}
        \caption{
            ZNE on the IBMQ Belem QPU experiment. The mitigated values obtained with the linear, quadratic, and exponential fittings were $-15.46551$, $-15.48897$, and $-15.49332$ Ha, respectively.
        }
        \label{fig:zne_qpu}
    \end{subfigure}

    \caption{
        Zero-noise extrapolation (ZNE) results from IBMQ Belem's noise model (a) and the real Belem QPU (b). Both are obtained using the \textit{Efficient SU2} ansatz and $4000$ shots per measurement. No other error mitigation was applied on the ZNE measurements.
    }
    \label{fig:zne_model_and_qpu}
\end{figure*}

%
%

In Fig. \ref{fig:zne_effsu2_model}, we show the results of quantum error mitigation using the ZNE technique applied to noisy \textit{Efficient SU2} ansatz results, using the IBMQ Belem noise model in the simulation. We used $4000$ shots to compute the energies at the three noise levels, $\{1, 3, 5\}$, used for ZNE, fitted with linear, quadratic, and exponential functions.  The procedure results in absolute errors, defined as $|E_0^\text{target} - E_0^\text{ZNE}|$, of $0.06600$, $0.04752$, and $0.04465$ Ha for the linear, quadratic, and exponential fittings respectively. The latter is more accurate than the simulated noisy energy by $0.02589$ Ha. The very same technique was performed on the IBMQ Belem QPU, as presented in Fig. \ref{fig:zne_qpu}, and yielded absolute errors of $0.09538$, $0.07192$, and $0.06757$ Ha for the same fittings. We note that the exponential fitting function is of the form: $a e^{bx}+c$.

\begin{table}[h!]
\centering
\begin{tabular}{lccc} 
    \hline
    Result & Energy (Ha) & $|\Delta E|$ & $\sigma$ \\[0.7ex] 
    \hline
    Noisy \textit{UCCSD}                  & -14.79180   & 0.76909   & 0.00860        \\[0.5ex]
    QPU \textit{Efficient SU2}            & -15.43317   & 0.12771   & 0.41725$^*$    \\[0.5ex]
    Noisy \textit{Efficient SU2}          & -15.49035   & 0.07054   & 0.00404        \\[0.5ex]
    QPU \textit{Efficient SU2} + ZNE      & -15.49332   & 0.06757   & -              \\[0.5ex]
    Noisy \textit{Efficient SU2} + ZNE    & -15.51624   & 0.04465   & -              \\[0.5ex]
    QPU \textit{Efficient SU2} + T-REx    & -15.53275   & 0.02814   & 0.46579$^*$    \\[0.5ex]
    Perfect \textit{Efficient SU2}        & -15.56043   & 0.00046   & 0.00008        \\[0.5ex]
    Perfect \textit{UCCSD}                & -15.56077   & 0.00012   & 0.00003        \\[0.5ex]
    Target energy                         & -15.56089   &  -        & -              \\[0.7ex] 
    \hline
\end{tabular}
\caption{Summary of results. The energy results for the noisy and perfect simulations are given as averages over the best $5$ VQEs of each experiment, along with the associated standard deviation $\sigma$. $|\Delta E|$ is the difference from the target energy. The ZNE-mitigated values correspond to the exponential fittings. $^*$\emph{The QPU's standard deviation is as given in Qiskit's Estimator results and corresponds to a single 4000-shot measurement}.}
\label{table:results}
\end{table}

As shown in Table \ref{table:results}, energy evaluation from the VQE implemented on the real quantum hardware results in a lower error than the noisy simulations results even when ZNE is applied. This highlights the efficiency of the T-REx mitigation technique when implemented on the QPU.

\section{Discussion}
\label{sec:discussion}
When implementing our VQE on ideal device, both ans\"atze ended up converging towards the target $E_0^{\text{target}}$ within less than $1$ millihartree when taking the average over the best $5$ VQE results, with the \textit{UCCSD} ansatz performing better than \textit{Efficient SU2}, as is shown in Table \ref{table:results}. Both ans\"atze also converged toward an energy value below the Hartree--Fock energy, $E_{\rm HF} = -15.56033$ Ha, with $\avg{E_\text{\rm Eff SU2}} = -15.56043$ Ha, and $\avg{E_\text{\rm UCCSD}} = -15.56077$ Ha. This means that the molecule's correlation energy was also captured by the VQE-found ground state, which is important since $E_{\rm HF}$ can be obtained using only single-gate rotations and dropping all entanglements. 
Moreover, we notice a faster convergence for the \textit{UCCSD} compared to \textit{Efficient SU2}, but with a few of VQEs ending up converging towards a local minimum situated at around $-15.25$ Ha.
In the presence of simulated noise, the two ans\"atze performed differently: the \textit{Efficient SU2} ansatz VQEs now converge towards a slightly higher energy value, $\avg{E_\text{Eff SU2}^{\rm noisy}} = -15.49035$ Ha, compared to the ideal case, whereas the \textit{UCCSD} VQEs have not converged at all, with $\avg{E_\text{UCCSD}^{\rm noisy}} = -14.79180$ Ha. The large convergence difference between the hardware-efficient and the unoptimized chemically-motivated ans\"atze is expected and is due to their significant difference in circuit depth.

Naturally, the \textit{Efficient SU2} ansatz was subsequently used for the QPU experiment. In this work we have used the now retired 5-qubit IBMQ Belem QPU, and to remain within the constraints of QPU runtime, only SPSA gradient measurements were performed until $250$ iterations were reached, then a final energy measurement was carried out. T-REx, a light error mitigation, was used in this experiment throughout the whole VQE optimization. Even though we have used an older generation noisy QPU, the addition of light error mitigation allowed for results that were better than those obtained during noisy simulations, where not all sources of noise could be simulated. Explicitly, the obtained energy value from the QPU experiment was $E^{\rm QPU}_0 = -15.53275$ Ha. However, the effects of noise were still significantly apparent in the standard deviation of that QPU result. The associated QPU measurement variance returned by the IBM Quantum job was $\text{Var}^{\rm QPU} = 0.21696 \text{ Ha}^2$, corresponding to $\sigma^{\rm QPU} = 0.46579$ Ha.

In terms of computational resources, the use of T-REx for error mitigation introduced a constant measurement overhead of $8192$ shots at each iteration. This effectively increased the number of shots necessary for the VQE without error mitigation by 15\%, and so it did not constitute a major increase in the computation time. This overhead could be reduced further by not performing the calibration at each iteration, and rather perform it on an as-needed basis depending on the QPU. This was however not implemented in this work due to software limitations at the time.

The effect of light error mitigation via T-REx is better highlighted when examining the parameters $\{\vtheta_i\}$ produced during the VQE. Fig. \ref{fig:belem-qpu-on-svs} shows the classical energy evaluation on SVS where we find that the VQE had produced state parameters that describe a chemically accurate ground state. In fact, at iteration $121$, the SVS-evaluated energy with the parameter vector $\vtheta_{121}$ was $E^{\rm SVS}(\vtheta_{\rm 121}^{\rm QPU}) = -15.55979$ Ha, within chemical accuracy and only $1.1$ mHa away from the target energy value of our system. This is again despite QPU measurements misestimating the actual energy values during the VQE by giving higher values.

\begin{figure}[t!]
	\includegraphics[width=0.95\linewidth]{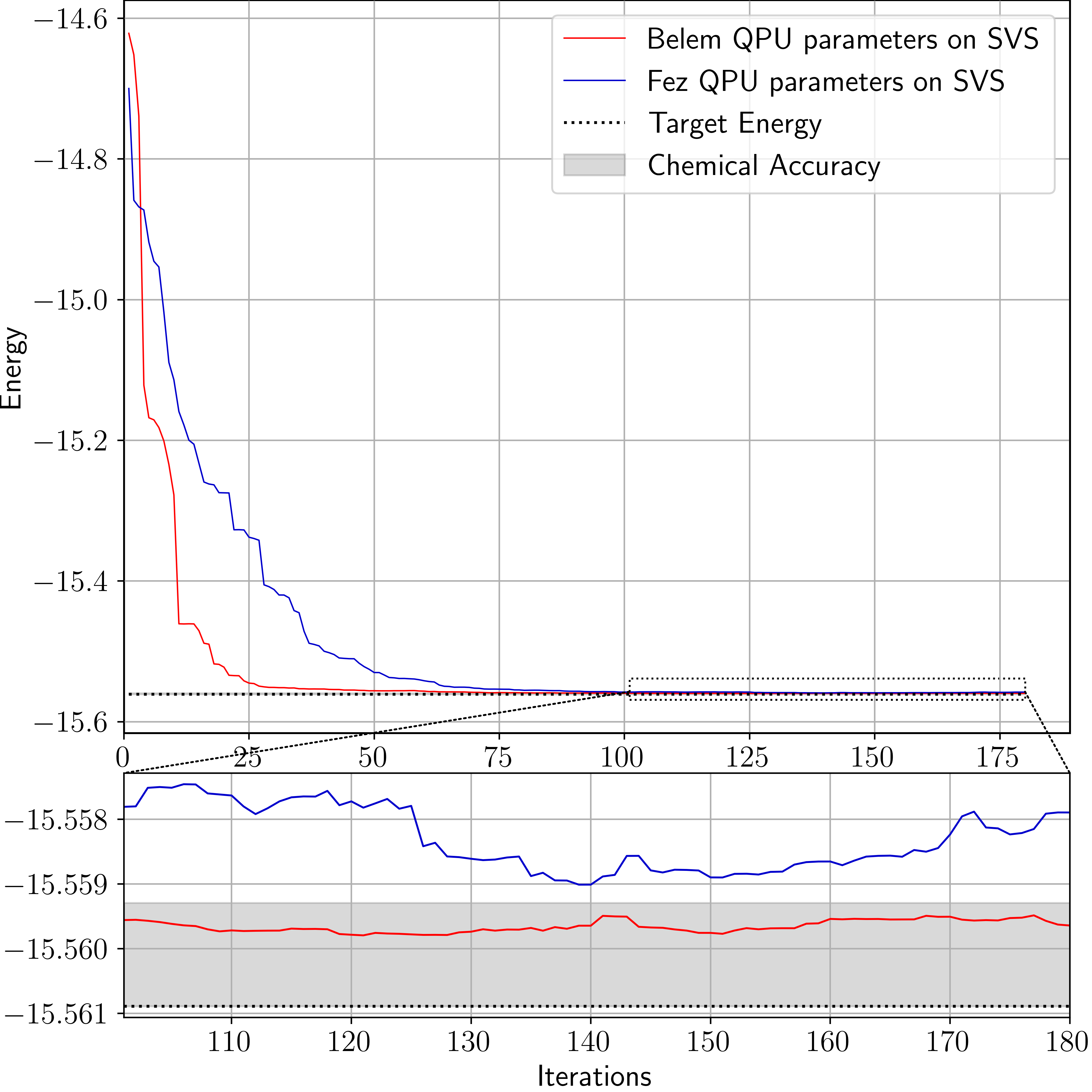}
	\centering
	\caption{Comparative graph of the SVS-evaluated energies resulting from this work (IBMQ Belem) versus those obtained using the more advanced IBM Fez obtained in \citet{Belaloui2025} without error mitigation. We only show 180 iterations as the IBM Fez experiment was cut off at that point.
    }
	\label{fig:belem_vs_fez_on_svs}
\end{figure}

The improvements resulting from using light error mitigation during the VQE optimization are further showcased when we compare our current results on IBMQ Belem with our previously published results \cite{Belaloui2025} obtained on the more recent and more advanced IBM Fez QPU, even though we are comparing two VQE runs on real quantum hardware due to accessibility limitations.
A comparative graph between the SVS-evaluated energies obtained in both experiments is given in Fig. \ref{fig:belem_vs_fez_on_svs}. It is worth noting that the comparison is carried out over the first 180 iterations, since the IBM Fez experiment was cut off at that 180 iterations limit. Indeed, comparing the ground-state results produced during both experiments by directly evaluating their resulting parameters classically shows that the VQE on IBMQ Belem, when supplemented with the light T-REx error mitigation, performed better---or at least no worse than the VQE that ran on IBM Fez, despite the gap in the QPUs' noise performance.

In the noise model simulations, when using the ZNE error mitigation technique with an exponential fitting function of the form $a e^{bx}+c$, we obtain a mitigated energy of $-15.51624$ Ha. Although the mitigated energy is still above the Hartree--Fock reference energy, it is closer to the target energy compared to the unmitigated noisy energy estimation, with $|\Delta E|^\text{noisy} = 0.07054$ Ha and $|\Delta E|^\text{noisy}_\text{ZNE} = 0.04465$ Ha. For the QPU case, the exponential fitting also gave the best result with an error $|\Delta E|^\text{QPU}_\text{ZNE} = 0.06757$ Ha, a significant improvement upon the unmitigated QPU energy measured in fold $1$ of the ZNE process. The latter gives an error $|\Delta E|^\text{QPU}_\text{raw} = 0.12771$ Ha, double the error compared to the ZNE-mitigated energy value. It is also worth noting that even though ZNE did significantly improve the energy measurement errors, especially in the QPU case where noise is stronger, the simpler T-REx mitigation method still produced a better QPU result. This is likely due to the presence of non-negligible coherent noise, which hindered ZNE from achieving better results. The optimal application of ZNE for QPU experiments in the presence of different levels of coherent noise is the subject of future work by the authors. This result, however, may suggest that for the shallow circuit we have implemented on QPU, readout errors were more significant than the errors induced by noise during the quantum computation.

\section{Conclusion}
\label{sec:conclusion}

This study demonstrates that, at least for small molecular systems which require a limited number of qubits, an older generation 5-qubit quantum processing unit (QPU), when augmented with light readout error mitigation, provides more accurate ground states and ground-state energy estimations than a more advanced 156-qubit device. This result shows the ability of error-mitigation strategies to enhance a QPU's practical capabilities, reflected here by compensating for the performance gap between QPUs of different generations.
When applied during the variational quantum eigensolver's (VQE) optimization loop, readout error mitigation (REM), and more specifically its optimized counterpart, the twirled readout error extinction (T-REx), significantly improves the accuracy of QPU results. Moreover, our results highlight the fact that the improvements are not only limited to the energy estimations, but also extend to the parameter optimization.\\

Indeed, the case for using light error mitigation is further strengthened when we compare the SVS-evaluated energies in our current results with the results in \citet{Belaloui2025}, effectively assessing the optimized variational parameters. Upon comparison between the two VQE runs, and although we have only considered a single VQE for each quantum computer due to cost limitations, the error-mitigated VQE running on IBMQ Belem notably performed better than the one ran on IBM Fez without error mitigation, as shown in Fig. \ref{fig:belem_vs_fez_on_svs}.
The application of T-REx not only refines the measurement of expectation values but, crucially, provides a more accurate and stable energy landscape within the VQE optimization loop, thereby guiding the classical optimizer to superior sets of variational parameters that more accurately describe the desired ground state.
It is worth mentioning that for practical applications such as this study's VQE, the computational cost of T-REx can be reduced further by optimizing its calibration step.\\

Additionally, further error mitigation such as zero-noise extrapolation (ZNE), when employed in post-processing, and eventually also in combination with lighter error mitigation schemes, refines even more the QPU energy estimation, yielding results that approach the target ground-state energy with greater accuracy. In applications where the actual ground-state energy value is the main result, and not merely the ground state itself, this heavier error mitigation in post-processing is essential in the noisy intermediate-scale quantum (NISQ) era. However, as this happens post-VQE, it would have no effect on the ground state itself, which is described rather by the optimized ansatz parameters. It should be noted that beyond heavy error mitigation, quantum methods such as the quantum-selected configuration interaction technique \cite{Kanno2023} can also be used to refine noisy VQE results and improve the accuracy of energy estimations.\\

Furthermore, this study confirms that greater emphasis should be placed on evaluating the quality of the variational parameters optimized by VQE, through classical methods where feasible, or via cost-efficient quantum-algorithmic alternatives, rather than focusing primarily on the final ground-state energy estimates. Several studies, such as \citet{Lively2024} and \citet{Skogh2024}, already underscore the primacy of higher quality parameters over the energy in the VQE's outcome for a variety of applications. This is particularly important as improving energy estimates tends to require the heaviest error mitigation strategies.\\

\section*{Acknowledgments}
This document has been produced with the financial assistance of the European Union (Grant no. DCI-PANAF/2020/420-028), through the African Research Initiative for Scientific Excellence (ARISE), pilot programme. ARISE is implemented by the African Academy of Sciences with support from the European Commission and the African Union Commission. The contents of this document are the sole responsibility of the author(s) and can under no circumstances be regarded as reflecting the position of the European Union, the African Academy of Sciences, and the African Union Commission.
We are grateful to the Algerian Ministry of Higher Education and Scientific Research and DGRST for the financial support.
We acknowledge the use of IBM Quantum services for this work. The views expressed are those of the authors, and do not reflect the official policy or position of IBM or the IBM Quantum team.\\

\bibliographystyle{unsrtnat}
\bibliography{refs.bib}

\end{document}